\documentstyle[11pt,aaspp4]{article}

\tighten
\eqsecnum

\begin{document}
%
\def\kms {km~s$^{-1}$}
\def\hkpc {$h^{-1}$kpc}
\def\hmpc {$h^{-1}$Mpc}
\def\kmsmpc {km~s$^{-1}$ Mpc$^{-1} \,$}
\def\lsim{ \lower .75ex \hbox{$\sim$} \llap{\raise .27ex \hbox{$<$}} }
\def\gsim{ \lower .75ex \hbox{$\sim$} \llap{\raise .27ex \hbox{$>$}} }
\def\items{\hangindent=0.5truecm \hangafter=1 \noindent}

\title{The Core Structure of Galaxy Clusters from Gravitational Lensing}

\author{Liliya L. R. Williams and Julio F. Navarro \altaffilmark{1}}
\affil{Department of Physics and Astronomy}
\affil{University of Victoria, Victoria, BC
V8P 1A1, Canada}

\and

\author{Matthias Bartelmann}
\affil{Max-Planck Institut f\"ur Astrophysik, Karl-Schwarzschild Strasse 1,\\
Garching bei M\"unchen, D-85740 Germany.}

\altaffiltext{1}{CIAR Scholar and Sloan Fellow}

\begin{abstract}
\noindent
We examine gravitational lensing constraints on the structure of galaxy clusters
and compare them with the results of cosmological N-body simulations of cluster
formation in cold dark matter (CDM) dominated universes. We find that cluster
core masses, as measured by the observed location of giant tangential arcs,
generally exceed those of dark matter halos of similar velocity dispersion. The
magnitude of the discrepancy is a strong function of cluster mass. Arc
properties in the most massive clusters in the sample (i.e. those with velocity
dispersion, $\sigma \sim 1500-2000$ \kms) are essentially consistent with the
N-body predictions.  On the other hand, giant arcs in $\sigma \sim \, 1000$
\kms \, clusters can only be reconciled with CDM cluster halos if their lensing
power has been increased substantially by the presence of a massive ($\sim \, 3
\times 10^{12} \, h^{-1} M_{\odot}$) central galaxy and of significant
substructure. Best agreement is found if the mass of the central galaxy and the
effects of substructure are approximately independent of cluster mass. Massive
central galaxies with steep inner density profiles are also needed to explain a
clear trend observed in our dataset between the radial thickness of giant
tangential arcs and the velocity dispersion of the cluster lens.  The position
and redshift of radial arcs may be used as independent tests of these results,
but at present the dataset available is too limited to have a significant impact
on these conclusions. Our results depend only weakly on the cosmological model
adopted, and suggest that structural parameters of clusters derived from strong
lensing studies cannot usefully constrain the values of cosmological parameters.
\end{abstract}

\section{Introduction}

The extraordinary lensing power of galaxy systems first put in evidence by the
discovery of giant arcs (Lynds \& Petrosian 1986, Soucail et al 1987a) provides
an invaluable tool for investigating the gravitational potential of galaxy
clusters at moderate redshift. For example, early studies of the giant arcs were
instrumental in cementing the view that a monolithic dark matter halo dominates
the cluster gravitational potential, to which individual galaxies contribute
relatively small local perturbations (Soucail et al 1987b).  Detailed studies of
the location, morphology, and magnification of giant arcs provide further
insight into the mass structure within clusters (AbdelSalam et al 1998, Kneib et
al 1996). 

The most straightforward interpretation of the location of giant tangential arcs
is that they occur roughly at the ``Einstein radius'' of a cluster, allowing
accurate estimates of the total projected mass enclosed within the arc if the
angular diameter distances to the lensing cluster and to the arc source galaxy
can be measured. This method remains to date the most direct estimator of the
total mass projected onto the core of a galaxy cluster.  Accurate estimates of
the core surface mass density within the arc may be used to place strong upper
limits on the core radii of isothermal cluster mass models, and early results
were puzzling. Narayan, Blandford \& Nityananda (1984) noted that the small
cores required to explain the properties of giant arcs were at odds with the
relatively large core radii derived from X-ray and optical observations. Small,
but finite, core radii were also required to account for observations of radial
arcs in clusters such as MS2137-23 (Fort et al 1992) and A370 (Smail et al
1995).

A simple explanation for this discrepancy was proposed by Navarro, Frenk \&
White (1996, hereafter NFW96) on the basis of cosmological N-body simulations of
cluster formation. These authors found that isothermal models are in general a
poor approximation to the structure of dark halos formed in N-body simulations,
and proposed a simple model to describe the structure of dark matter halos. In
this model, which we shall refer to as the ``NFW'' model, the density profile is
shallower than a singular isothermal sphere near the center, and steepens
gradually outwards to become steeper than isothermal far from the center. This
result explains naturally why models based on the isothermal sphere fail to
account simultaneously for lensing and X-ray observations. X-ray core radii,
which correspond to the radius where the {\it local slope} of the mass profile
is close to the isothermal value, occur in NFW halos at different radii than
giant arcs, whose location trace the radius where the {\it average inner surface
density} equals the ``critical'' lensing value (see eq.~9 below).

Subsequent work by Navarro, Frenk \& White (1997, hereafter NFW97) showed that
the structure of dark halos is approximately independent of mass, power
spectrum, and cosmological parameters, and demonstrated how a simple algorithm
may be used to calculate the mass profile of halos of arbitrary mass in
hierarchical universes. The procedure applies only to halos that are close to
dynamical equilibrium and assumes spherical symmetry, but it has no free
parameters and can be used to predict the location and magnification of giant
tangential arcs (as well as of radial arcs) once the velocity dispersion of the
cluster and the angular diameter distances to the source galaxy and the cluster
lens are specified. Thus, in principle, lensing observations may be used to test
directly the overall applicability of the results of N-body simulations to the
structure of dark halos on the scale of galaxy clusters.

In this paper we compile the results of gravitational lensing studies of 24
galaxy clusters to investigate whether the properties of gravitationally lensed
arcs are consistent with the NFW dark halo model. We discuss the lensing
properties of NFW halos in \S2 and summarize the main properties of our dataset
in \S3. Section 4 presents our main results. In \S5 we discuss the implications 
of our model, and in \S6 we summarize our main findings.

\section{Lensing Properties of NFW halos}

\subsection{The NFW mass profile}

As discussed by NFW96 and NFW97, the spherically averaged density profiles of
equilibrium dark matter halos formed in cosmological N-body simulations of
hierarchically clustering universes are well represented by a simple formula,
$$
{\rho(r) \over \rho_{crit}}={\delta_c \over (r/r_s)(1+r/r_s)^2}, \eqno(1)
$$
where $\rho_{crit}=3H^2/8 \pi G$ is the critical density for closure, $H(z)$ is
the Hubble parameter,\footnote{We parameterize the present
value of Hubble's constant by $H_0=100 \, h$ \kms Mpc$^{-1}$.} $\delta_c$ is a
dimensionless characteristic density contrast, and $r_s$ is a scale radius. If
we define the mass of a halo, $M_{200}$, as the total mass of a sphere of mean
density $200$ times critical, eq.(1) above has a single free parameter once the
halo mass is specified. (The radius of this sphere, $r_{200}$, is sometimes
called the ``virial'' radius.) The free parameter can be taken to be the
characteristic density contrast $\delta_c$ or the ``concentration'' parameter,
$c=r_{200}/r_s$. These two parameters are related by
$$
\delta_c={200 \over c} {c^3 \over \ln(1+c)-c/(1+c)}. \eqno(2)
$$
The numerical experiments of NFW97 indicate that $\delta_c$ is determined by the
mean matter density of the universe at the redshift of collapse of each halo,
i.e.  $\delta_c(M_{200}) \propto (1+z_{coll}(M_{200}))^3$. Halos of increasing
mass collapse later in hierarchical universes, and therefore $\delta_c$ and $c$
are monotonically decreasing functions of $M_{200}$. Collapse redshifts can be
easily calculated once the cosmological model is specified, and NFW97 describe a
simple algorithm that can be used to calculate the density profile of halos of
arbitrary mass in cold dark matter dominated universes (see the Appendix of
NFW97 for details).

\subsection{Lensing by NFW halos}

The lensing properties of axially symmetric lenses is described in detail in
Schneider, Ehlers \& Falco (1992) and in the many reviews of gravitational
lensing (see, e.g., Blandford \& Narayan 1992, Narayan \& Bartelmann
1995). Provided that the gravitational potential causing the deflection is
small, $|\Phi| \ll c^2$, the lens equation describing the mapping of the source
plane into the image plane is very simple. In terms of the angular diameter
distances to the lens ($D_l$), to the source ($D_s$), and between lens and
source ($D_{ls}$), a lens is locally described by the Jacobian matrix of the
mapping,
$$
A=\left(\delta_{ij}-{\partial^2 \psi \over \partial \theta_i \partial
\theta_j}\right), \eqno(3)
$$
where ${\vec \theta}=(\theta_i,\theta_j)$ are angular coordinates relative to
the optical axis, and $\psi$ is the projected Newtonian potential of the lens,
$$
\psi({\vec \theta})={D_{ls} \over D_l D_s} {2 \over c^2} \int_{-\infty}^{+\infty}
\Phi(D_l{\vec \theta},z) dz \eqno(4)
$$
The lensing properties of NFW halo models are fully specified by the radial
dependence of the surface mass density in units of the critical surface mass
density $\Sigma_{cr}$; the ``convergence'',
$$
\kappa(x)={\Sigma(x) \over \Sigma_{cr}}, \eqno(5)
$$
where $\Sigma_{cr}$ depends on the lens-source configuration through
$$
\Sigma_{cr}={c^2 \over 4 \pi G} {D_s \over D_l D_{ls}}, \eqno(6)
$$
and $x=r/r_s$ is the radius in units of the NFW scale radius. Following the
derivation of Bartelmann (1996), the mass inside radius $x$ can be described by
the dimensionless function,
$$
m(x)=2 \int_0^{x}\kappa(y) y dy, \eqno(7)
$$
which can be used to find the eigenvalues of the Jacobian matrix $A$,
$$
\lambda_r=1-{d \over dx}{m \over x} \eqno(8)
$$
$$
\lambda_t=1-{m \over x^2}. \eqno(9)
$$
The tangential and radial critical curves occur at $x_t$ and $x_r$, where
$\lambda_t=0$ and $\lambda_r=0$, respectively. The surface density associated
with an NFW model (eq.1) is
$$
\Sigma(x)={2 \, \delta_c \, \rho_{crit} \, r_s \over x^2-1} f(x), \eqno(10)
$$
with
$$
f(x)= 
\cases{
{1-(2/\sqrt{x^2-1}) \tan^{-1} \sqrt{(x-1)/(x+1)}} & if $x > 1$;\cr
{1-(2/\sqrt{1-x^2}) \tanh^{-1} \sqrt{(1-x)/(1+x)}} & if $x < 1$;\cr
  0 & if x=1.\cr} \eqno(11)
$$
Defining $\kappa_s=\delta_c \, \rho_{crit} \, r_s/\Sigma_{cr}$, we can write the
convergence as
$$
\kappa(x)={2 \, \kappa_s \over x^2-1} f(x), \eqno(12)
$$
and the dimensionless mass $m(x)$ as,
$$
m(x)=4 \, \kappa_s \, g(x), \eqno(13)
$$
with
$$
g(x)=\ln{x \over 2} +
\cases{
{(2/\sqrt{x^2-1}) \tan^{-1} \sqrt{(x-1)/(x+1)}} & if $x > 1$;\cr
{2/\sqrt{1-x^2}) \tanh^{-1} \sqrt{(1-x)/(1+x)}} & if $x < 1$;\cr
  1 & if x=1.\cr} \eqno(14)
$$
Equations (10)-(14), together with the algorithm to compute halo parameters
outlined by NFW97, can be used to estimate the location of tangential and radial
arcs for halos of arbitrary mass.

Finally, we note that the arc thickness is controlled by the angular size of the
source and by the value of the convergence at the critical line. As demonstrated
by Kovner (1989) and Hammer (1991), the thin dimension of an arc is magnified by
a factor of order $\mu \approx 1/2(1-\kappa)$ relative to its original angular
size. Tangential arcs thinner than the source thus require $\kappa(x_t)<0.5$. We
use these results below to analyze the constraints posed by observations of
giant arcs on the structure of galaxy clusters.

\section{The Dataset}

The main properties of clusters in our sample are listed in Table 1. The sample
includes all systems in the recent compilation by Wu et al (1998) with measured
velocity dispersion. The table lists the following information for each cluster:
(1) cluster name, (2) redshift, (3) velocity dispersion in \kms, (4) designation
of the arc used in the analysis (as labeled in the appropriate reference), (5)
the redshift of the arc (if available), (6) the arc distance from the center of
the cluster, typically chosen to coincide with that of the brightest cluster
galaxy, in arcsec, (7) the radial half-light thickness of the arc, in arcsec
(upper limits correspond to the seeing of the observation when arc is
unresolved), (8) telescope and instrument used, and (9) references for the arc
width.  

Some values in Table 1 differ from those adopted by Wu et al (1998). {\bf A370:}
The giant arc in this strongly bimodal cluster is 10'' from the nearest bright
galaxy and 26'' from the center of mass of the cluster. As a compromise we take
the clustercentric distance of the arc to be 18''.  {\bf AC114:} The most
prominent arc, A0, is almost certainly a singly imaged source located beyond the
cluster's critical curve. The location of the critical curve is thus not well
determined but must lie somewhere inside that radius. There is a multiply imaged
system in the same cluster, $S1/S2$. We estimate the clustercentric distance of
the tangential critical line to be the average of A0 and $S1/S2$, or 38''. {\bf
MS0440:} We take the velocity dispersion of this cluster to be 872 \kms (Gioia et al
1998) rather than 606 \kms (Carlberg et al. 1996), because the former value is more
consistent with the cluster's temperature, $T_X=5.5$ keV, and its (0.1-2.4) keV
X-ray luminosity, $L_X=7.125\times10^{43}h^{-2}$ erg s$^{-1}$. {\bf Cl0024:} The
redshift of the source galaxy has been determined spectroscopically, $z_s
\approx 1.7$ (Broadhurst et al. 1999). {\bf A2124:} Data for this cluster have
been taken from Blakeslee \& Metzger (1998). {\bf Cl0016+1609:} Data for this
cluster have been taken from Lavery (1996).

The column labeled $\delta\theta_t$ in Table 1 lists the half-light radii of
the giant arcs in the radial direction (or the half-seeing if the arc is unresolved).
In clusters where more than one tangential arc has width information, the
average of the available widths is listed. Except for Cl0302, where Mathez et al
(1992) measure an arc half-width of 0.6'' for the A1/A1W pair and Luppino et al
(1999) quote $<0.25$'' for the same arc, width estimates from different authors
agree to within the errors in all clusters.  We take the average of the two
discrepant values for Cl0302.

\section{Results}

\subsection{Main trends in the dataset}

Figures 1 and 2 summarize the main trends in the dataset. The top panel in
Figure 1 shows the velocity dispersion of the cluster ($\sigma$) versus the
clustercentric distance of the giant tangential arc ($\theta_t$). Assuming
circular symmetry, arc distances can also be expressed in terms of the total
projected mass within the arc radius, $M_{core}=\pi \, \Sigma_{cr} \, (\theta_t
D_l)^2$. This mass estimate depends, through $\Sigma_{cr}$ and $D_l$, on the
angular diameter distances to source and lens. We have assumed a simple
Einstein-de Sitter cosmological model to compute the values of $M_{core}$
plotted in the bottom panel of Figure 1. Arcs without measured redshifts are
assumed to be at $z_t=1${\footnote{We shall hereafter use subscripts $t$ and $r$
to refer to quantities associated with the source of tangential and radial arcs,
respectively.}}.

The first thing to note from Figure 1 is that core mass and lensing power seem
to correlate only weakly with velocity dispersion in clusters with $\sigma \gsim
\, 1000$ \kms. This is at odds with scalings expected from simple
models. For comparison, the dotted lines in these panels indicate the Einstein
radius and the core mass of a singular isothermal sphere at $z_l=0.3$ lensing a
source galaxy located at $z_t=1.0$ in an Einstein de Sitter universe. The strong
dependence on $\sigma$ expected in this simple model ($M_{core} \propto
\sigma^4$) is clearly absent in the data. We note as well that several clusters are
more powerful lenses than singular isothermal spheres, indicating a large
central concentration of mass in some of these systems. 

The filled circles in Figure 2 indicate the (half-light) radial widths of the
giant tangential arcs quoted in the literature (see Table 1 for
references). Open circles are upper limits to the width derived from the seeing
at the time of observation in cases where the arc is unresolved. A clear trend
is observed between width and cluster velocity dispersion: arcs become thicker
as $\sigma$ increases. The trend is highly significant. Treating upper limits as
measurements, $96.9\%$ of randomly reshuffled ($\sigma$, $\delta\theta_t$)
samples have a smaller Kendall $\tau$ correlation coefficient than the real
sample.  Neglecting the one deviant point (which corresponds to the arc in
Cl0016, deemed unresolved in an HST WFPC2 image by Lavery 1996) increases the
significance of the correlation as measured by this test to $99.1\%$.
Because upper limit points are confined exclusively to the low-$\sigma$ section
of the diagram, our procedure most likely underestimates the significance of the
correlation.

As discussed in \S2.2, the ratio ($\mu_r$) between the radial thickness of
tangential arcs and the intrinsic angular extent of the source depends directly
on the value of the convergence at the location of the arc,
$\mu_r=1/2[\kappa(x_t)-1]$. Within this context, the correlation shown in Figure
2 implies that the convergence at the arc location, $\kappa(x_t)$, increases
systematically with $\sigma$. The actual values of $\kappa(x_t)$ depend on the
intrinsic angular size and redshift of the source, which we shall now examine.
From Table 1, many of the sources with measured redshifts are at $z_t\sim 0.7$.
We compare in Figure 3 their ``lensed half-widths'' $\delta \theta_t$ with the
angular size of field galaxies at various redshifts: (i) the half-light radii of
galaxies at intermediate redshifts in the WFPC Medium Deep Survey (Mutz et al
1994), (ii) the half-light radii of $z\sim 1$ galaxies in the CFRS sample of
Lilly et al (1998), and (iii) the half-light radii of ``Ly-break'' galaxies at
$z \approx 3$ (starred symbols in Figure 3, from Giavalisco et al 1996).

The data in Figure 3 imply that the angular size of galaxies decreases
monotonically out to $z\sim 3$. Taken at face value, intrinsic galaxy sizes
also seem to decrease as a function of $z$, in agreement with predictions from
hierarchical models of galaxy formation (Mo, Mao \& White 1998). One crude
estimate of the evolution may be made by comparing the observational data with
the angular extent of a ``standard rod'' of fixed proper size equal to the
average half-light radius of bright ($\sim L_{\star}$) spirals, $\approx 4.4
h^{-1}$ kpc (top set of curves, Mutz et al. 1994),
and with the angular radius of sources whose proper size scale in inverse
proportion to $(1+z)$ (bottom set of lines). The actual evolution in source size
out to $z\sim 1$ is approximately intermediate between these two somewhat
extreme examples. A word of caution applies to this conclusion. The three
surveys shown in Figure 3 have been conducted in different passbands, are
subject to different selection biases, and may sample intrinsically different
source populations. For example, the Ly-break galaxies seem to be forming stars
preferentially in their central regions and therefore would appear substantially
smaller in the rest-frame UV used to estimate their sizes than the half-light
radii of more passively evolving galaxies examined by the other groups.

With this caveat, we observe that arcs with measured redshifts (filled circles
in Figure 3) seem to be of angular size comparable to, or thinner than, field
galaxies at similar redshifts.  In particular, half-light radii of galaxies in
the CFRS survey exceed the arcwidths by about $50 \%$. This may be due in part
to the fact that the magnitude-limited CFRS sample is biased towards the bright,
large galaxies present at that redshift, while heavily magnified arc sources may
be intrinsically fainter and smaller. We conclude, rather conservatively, that
the radial magnification of giant tangential arcs probably does not exceed
unity, $\mu_r \, \lsim \, 1$, implying that $\kappa(x_t) \lsim 0.5$.

\subsection{Interpretation of the observed trends}

The trends highlighted in Figures 1 and 2 and, in particular, the correlation
(or lack thereof) between $\sigma$ and $\theta_t$ suggest that the lensing
properties of galaxy clusters differ significantly from those of simple
circularly symmetric models such as NFW or the singular isothermal sphere. A
number of effects may be responsible for the disagreement; e.g., asphericity in
the mass distribution, uncertainties in velocity dispersion estimates,
substructure associated with departures from dynamical equilibrium, and the
presence of a massive central galaxy.

Let us consider first the effects of asphericity in the mass
distribution. N-body models suggest that the distribution of mass in equilibrium
galaxy clusters deviates significantly from spherical symmetry, and is well
approximated by triaxial shapes maintained by anisotropic velocity dispersion
tensors (Thomas et al 1998 and references therein). Estimates of the cluster
velocity dispersion and giant arc properties are therefore sensitive to the
relative orientation between the principal axes and the line of sight to the
cluster. For example, line-of-sight velocity dispersion estimates of
cigar-shaped clusters observed end-on would lead to systematic overestimation of
the true average $\sigma$, but $\theta_t$ would be similarly affected by the
favorable orientation, in a manner that preserves a firm correlation between
$\sigma$ and $\theta_t$.

Another factor that may affect the observed relation between $\sigma$ and
$\theta_t$ are observational uncertainties in $\sigma$ estimates. Velocity
dispersions are notoriously difficult to estimate properly, as the error budget
is generally dominated by systematic uncertainties such as cluster membership
rather than by strict measurement error (Zabludoff et al 1990, Carlberg et al
1996). The magnitude of the errors required to cause the apparent lack of
correlation between $\sigma$ and $\theta_t$ (of order $1000$
\kms) appears, however, excessive. We conclude that projection effects and
observational error may contribute significantly to the {\it scatter} in
correlations between lensing and dynamical properties but are unlikely to be the
source of the trends shown in Figures 1 and 2.

A simpler alternative is that the projected surface density profile inside
$\theta_t$ effectively steepens towards lower $\sigma$.  Indeed, the effective
slope of the lensing potential can be deduced in a simple model-independent way,
based entirely on observables. Let us approximate the convergence {\it inside}
the tangential arc by a single power law, $\kappa(R)=\kappa_0 R^\alpha$, where
$\kappa_0$ and $\alpha$ are functions of $\sigma$ and $R$ is the projected
radius. Applying the condition that inside the location of the tangential arc the
average value of the convergence is unity, and that the value of $\kappa(R_t)$
at the location of the arc is known from arc's width magnification,
we derive the relation
$\alpha=2[\kappa(R_t)-1]$. In other words, the effective slope of the projected
density profile is inversely proportional to the width magnification of the
arc. From the data presented in Figures 1 and 2 we see that a $\sigma \sim 1000$
\kms cluster has $\kappa(R_t) \approx 0.35$ and $\alpha\approx -1.3$. On the
other hand, $\sigma_v\, \gsim \, 1500$ \kms clusters have much shallower
profiles; $\kappa(R_t)\approx 0.6$ and $\alpha\approx -0.8$.  {\it The tangential arc
properties of clusters in our sample imply that the effective lensing potential
is steeper (shallower) than isothermal in $\sigma \, \lsim \, 1250$ ($\sigma \,
\gsim \, 1250$) \kms clusters.} (A singular isothermal sphere has $\alpha=-1$ in
this notation.)

As noted in \S4.1, the conclusion that the slope of the inner density profile is
a strong function of cluster mass is difficult to reconcile with the nearly
scalefree structure of cold dark matter halos found in cosmological N-body
simulations. This discrepancy afflicts all models where the core structure of
dark halos is approximately independent of mass. In particular, slight
modifications to the NFW profile that preserve its independence of scale, such
as those proposed by Moore et al (1998) and Kravtsov et al (1998), would also
fail to reproduce the lensing observations.

We investigate below whether the lensing data may be reconciled with scalefree
models such as NFW by assuming that the lensing power of clusters has been
significantly boosted by the presence of dynamical substructure and of massive
central galaxies, in a way that mimics the correlations between effective slope
and $\sigma$ pointed out above. We emphasize that qualitatively our conclusion
applies to all scalefree models of halo structure but we adopt below the NFW
description in order to derive {\it quantitative} estimates of the lensing
contribution by the central galaxy and by substructure.

\subsection{Comparison with NFW halo models}

As discussed in \S2, the lensing properties of NFW halos can be computed as a
function of velocity dispersion once the lens-source configuration is
specified. This allows us to compare the lensing data directly with the
predictions of the model. The comparison depends on the values of the
cosmological parameters, since these control both the halo parameters (NFW97)
and the angular diameter distances of each lens-source
configuration. Qualitatively, however, none of the conclusions we discuss below
depend on this choice of cosmological model. For illustration, we explore first
the lensing properties of NFW clusters in a CDM-dominated universe with
$\Omega_0=0.2$, $\Lambda=0.8$, and $h=0.7$. The power spectrum has been
normalized by $\sigma_8=1.13$ in order to match the present day abundance of
galaxy clusters, as prescribed by Eke, Cole \& Frenk (1996). As in Figure 1, we
assume that arcs without measured redshift are located at $z_t=1$.

The top-left panel in Figure 4 shows the ratio between the observed and
predicted ``Einstein radius'' as a function of cluster velocity
dispersion. Clearly, CDM halos formed in this cosmology are in general less
powerful lenses than actual clusters. This result is not unexpected given our
previous discussion: most $\sigma \sim 1000$ \kms \, lenses are more powerful
than singular isothermal spheres, let alone models with shallower inner density
profiles such as NFW. The magnitude of the discrepancy depends strongly on
$\sigma$. Clusters with $\sigma\sim 1000$ \kms \, have Einstein radii about a
factor of three larger than expected for NFW models. On the other hand, the
light-deflecting power of the most massive clusters in the sample (i.e. those
with $\sigma \sim 1500-2000$ \kms) agrees well with that of NFW models. (We use
for the comparison the halo mean velocity dispersion within the virial radius,
$\sigma_{200}=(GM_{200}/2 \, r_{200})^{1/2}$.)

A compounding problem that afflicts cluster mass models with shallow inner
density profiles such as NFW was noted by Bartelmann (1996) and concerns the
radial magnification of tangential arcs: the convergence at the tangential
critical curve is $\kappa(x_t)>0.5$, implying that the radial magnification
exceeds unity. This is shown by the empty circles in the top-left panel of
Figure 5. A crude estimate of the ``true'' radial magnification can be made for
arcs with measured redshifts by assuming that the actual source half light radii
is $4.4 (1+z_t)^{-1/2} \, h^{-1}$ kpc (see Figure 3). Values of $\mu_r$ computed
this way are shown as thick crosses in Figure 5. The difficulty pointed out by
Bartelmann (1996) is confirmed by our analysis: observed tangential arcs are
roughly $2$-$3$ times thinner than expected from NFW model lenses.

\subsection{Dependence on the cosmological parameters}  

Strictly speaking, the quantitative estimates of the disagreement between NFW
models and observations presented above are valid only for the low-density
$\Lambda$CDM model adopted there, but qualitatively the conclusions are
independent of the cosmological parameters. For example, assuming CDM universe
models normalized to match the present day abundance of galaxy clusters, we find
that changing $\Omega_0$ from $0.2$ to $1$ (in flat or open geometries) modifies
$\theta_{t}/\theta_{t,pred}$ and $\mu_r$ only by about $10$-$20\%$, a negligible
effect compared to the effects of central galaxy and substructure we explore
below.

\subsection{The role of substructure}

As discussed by Bartelmann, Steinmetz \& Weiss (1995), the discrepancy shown in
Figure 4 between observed and predicted tangential arc clustercentric distances
may be ameliorated by considering the effects of substructure. Let us
parameterize this effect by the outward displacement of the tangential arc
relative to a circularly averaged NFW halo of the same velocity dispersion,
$f_t=\theta_{t}/\theta_{t,{\rm NFW}}$. The distribution of $f_t$ has been
determined from N-body simulations (Bartelmann \& Steinmetz 1996) and is shown
by the solid histogram in Figure 6; on average substructure pushes out
tangential arcs by $50\%$ in radius. Since there is little indication either
from observations or numerical simulations that substructure is a strong
function of mass on the scales we probe here, we will assume that all clusters
are affected equally, regardless of $\sigma$. The upper right panels of Figures
4 and 5 indicate the result of increasing $\theta_{t,pred}$ by $\sim 50\%$ in
order to account for this effect. The error bars correspond to the $1/4$ and
$3/4$ quartiles in the distribution of $f_t$ (Figure 6). Note that width
magnifications are also reduced because the convergence at the tangential
critical curve decreases as the arc moves outwards (Figure 5). Including
substructure helps to narrow the gap, but it appears insufficient to reconcile
fully the predictions of NFW halo models with observations.

\subsection{The effects of a central massive galaxy}

The interpretation advanced in \S4.2 ascribes the remainder of the difference to
the lensing contribution of a central massive galaxy, an ansatz that allows us
to estimate its total projected mass within the tangential arc radius. This mass
may be compared directly with the stellar mass in the central galaxy, which we
estimate as follows. The typical absolute magnitude of brightest cluster
galaxies is $M_V-5\log{h}\approx -23.5$ (Schombert 1986), which combined with
rough upper limits to the stellar mass-to-light derived from lensing of QSOs by
isolated ellipticals, $M/L_V \approx 15 h M_{\odot}/L_{\odot}$ (Keeton, Kochanek
\& Falco 1998), imply a total stellar mass of order $M_g \approx 3 \times 10^{11}
\, h^{-1} \, M_{\odot}$. Assuming that the galaxy mass profile is well approximated by
a de Vaucouleurs law with effective radius, $r_{eff}=15 \, h^{-1}\,$ kpc, we
recompute the predicted location of tangential arcs including this contribution
and report the results in the bottom-left panels of Figures 4 and 5. The results
assume that the structure of the dark halo is modified ``adiabatically'' by the
presence of the central galaxy (see details in NFW96) and are insensitive to our
choice of effective radius provided that $r_{eff}\ll \theta_t D_l\approx
60$-$120 h^{-1}$ kpc; i.e., provided that most of the galaxy's mass is contained
within the tangential critical line. Most brightest cluster galaxies studied so
far easily meet this criterion (Schombert 1986).

Predicted arc locations in the bottom-left panel of Figure 4 include the
combined effects of substructure and of the stellar component of the central
cluster galaxy but are still seen to fall short of observations. This result is
rather insensitive to the choice of cosmological model. The disagreement
actually grows more acute if higher density universes are considered because
$\Sigma_{cr}$ and, therefore lens masses, increase monotonically with $\Omega_0$
for the lens-source configuration we assume here. Good agreement with
observations require that the total mass associated with the central galaxy is
increased significantly over and above the expected stellar mass of the
galaxy. This is shown in the bottom-right panel of Figures 4 and 5, where we
have assumed that the total mass associated with the central galaxy is $M_g=3
\times 10^{12} h^{-1} M_{\odot}$, so that the average ratio of $\theta_t$
to $\theta_{t,pred}$ is about unity. This galaxy mass corresponds roughly to
the mass (projected inside $\theta_t$) of an isothermal sphere with velocity
dispersion of order $300$ \kms. This does not seem extravagant: velocity
dispersions measured for the central galaxies in MS1358+62, MS2053-04, and
MS2124 are all of order $\sim 300$ \kms \, or higher (Kelson et al 1997,
Blakeslee \& Metzger 1998), while in cluster RXJ1347.5-1145 the velocity
dispersion of stars in the central galaxy is $\sigma=620$ \kms (Sahu et
al. 1998). 

Reconciling lensing observations with NFW halo models thus require that the
central galaxy has somehow managed to retain a sizeable fraction of its own dark
halo. This may at first seem puzzling, but is consistent with observational
estimates of the mass attached to individual galaxies in clusters. For example,
based on the smooth structure of the arc in A370, Kneib et al (1993) conclude
that as much as $\sim 10^{11} h^{-1} M_{\odot}$ may be associated with
$M_B-5\log(h)\approx -19.6$ galaxies in that cluster. Brightest cluster galaxies
are about $\sim 30$ times more luminous and a simple scaling suggests that halos
as massive as $3\times 10^{12} h^{-1} M_{\odot}$ may indeed be associated with
central cluster galaxies. Our estimate is also consistent with the recent work
of Tyson, Fischer \& Dell'Antonio (1999), who argue that mass concentrations
surrounding individual galaxies in Cl0024 may be as large as $5 \times 10^{12}
h^{-1} M_{\odot}$. A detailed lensing study of A2218 by Kneib et al. (1996) also
indicates that individual cluster galaxies with velocity dispersion of order
$\sigma=300$ \kms must be surrounded with halos of mass $\sim
10^{12}\, h^{-1}M_\odot$.
We note as well that the total mass associated with the central galaxy would
actually be lower if, as proposed by Moore et al (1998), our procedure had
somehow underestimated the central density concentration of the cluster
halo. These authors report that NFW concentration parameter obtained from the
procedure outlined in NFW97 may be as much as $50\%$ lower than required to fit
their numerical experiments. Our conclusion does not change qualitatively, but
the mass of the central galaxy in this case would need to be revised downward by
approximately $50\%$ in order to fit the data.

\section{Implications of the model}

\subsection{Radial Arcs}

Our conclusion that the mass associated with the luminous central galaxy plays a
88888888fundamental role in the lensing properties of the cores of galaxy clusters can
be tested directly by considering the formation of radial arcs. These arcs are
located at the radial critical lines, $x_r$, where the eigenvalue $\lambda_r$
vanishes. We see from eqs.~8 and 9 that the location of the radial arc depends
on the angular gradient of the projected mass, rather than on the mean enclosed
surface density. Their location, therefore, may in principle be used to verify
independently our conclusion that the surface density profile steepens
systematically towards decreasing $\sigma$.

In the absence of a massive central galaxy, the relative location of radial and
tangential arcs formed through lensing by NFW halos is straightforward to
compute once the angular diameter distances to the sources are
known. Conversely, knowledge of the relative location of the arcs and of the
tangential arc redshift uniquely determines the redshift of the radial arc. For
example, Bartelmann (1996) applies this procedure to the radial/tangential arc
system in MS2137 and concludes that the sources of both arcs must be either at
very similar redshifts, or else far behind the cluster at $z \gg 1$. The
dichotomy simply reflects the fact that neither arc has measured redshift. The
tangential arc redshift is known for A370 (Soucail et al 1988), and the same
analysis yields a prediction $z_r \sim 1.5$ for the radial arc, in reasonable
agreement with the $z_r \approx 1.3$ prediction from the detailed lens models of
Kneib et al (1993) and Smail et al (1995). Radial arcs have now been observed in
four clusters: MS0440, MS2137, A370, and AC114. These clusters span the entire
range in velocity dispersion of our sample (AC114 has one of the highest
velocity dispersions, $\sigma=1649$ \kms, and MS0440 has one of the lowest,
$\sigma=872$ \kms), and therefore we expect that the systematic trends inferred
in the previous subsection may have a detectable influence on the properties of
the radial arcs.

One simple trend predicted by our interpretation concerns the relative location
of radial and tangential arcs as a function of the cluster velocity
dispersion. Assuming that the sources are at similar redshifts, the ratio
between the clustercentric distances of radial and tangential arcs,
$\theta_r/\theta_t$, depends strongly on the slope of the surface density
profile inside $\theta_t$: the steeper the profile the closer to the center the
radial arc moves and the smaller $\theta_r/\theta_t$ becomes. This is shown in
Figure 7, where the open circles represent the predictions of our model for all
clusters in our sample (including the effect of a central
galaxy of mass $M_g=3\times 10^{12} h^{-1} M_{\odot}$). We assume that the
redshifts of both arcs are the same, $z_r=z_t$, and adopt the same fiducial
cosmology of Figures 4 and 5. The arcs in MS0440, MS2137 and AC114 follow the
predicted trend very well, but the radial arc in A370 is much farther than
expected in our simple model. This may be because the source of the radial arc
is much farther behind the cluster than the tangential arc source; 
$z_r \approx 2 z_t \sim 1.4$, 
again in good agreement with the predictions of Smail et al (1995).

In summary, the relative location of radial and tangential arcs is a useful test
of the conclusions reached in the previous subsection regarding the role played
by the central galaxy on the lensing properties of clusters. Although there are
no measured redshifts for radial arcs, our analysis predicts that the arc
sources in MS0440, MS2137, and AC114 are at very similar redshifts. Consistency
with our model requires that the radial arc source must be far behind the
tangential arc in A370. Spectroscopic redshifts of radial arcs may thus be used
to verify or rule out the applicability to actual clusters of the mass modeling
we propose here.

\subsection{Observability of lensed features}

A second corollary of the interpretation outlined in \S4.2 is that, because
lensing features depend so heavily on the mass of the central galaxy, at fixed
cluster velocity dispersion tangential arc distances must correlate with the
total luminosity of the central galaxy. This could in principle be demonstrated
by comparing the residuals of the $\sigma$-$\theta_t$ correlation shown in
Figure 1 with residuals of the correlation between $\sigma$ and the total
luminosity of the central galaxy. A search of the literature yields,
unfortunately, few total absolute magnitudes for brightest cluster galaxies in
our sample, and therefore we are unable to utilize this test to verify our
interpretation. We intend to use archival images of clusters from different
telescopes and revisit this question in a future paper.

We note here that, if our interpretation is correct, it would help to explain
the apparent underrepresentation of moderate-$\sigma$ clusters in current lensing 
samples. Estimates based on the Press-Schechter (1974) algorithm predict that
clusters with $\sigma \gsim 1000$ \kms \, should outnumber clusters with $\sigma
\gsim 1500$ \kms \, by a factor of $\sim 20$ or more (Eke et al 1998), a result of
the exponential decline of the cluster mass function at the high-mass end. This
sharp decrease in the number of clusters with increasing $\sigma$ is not readily
apparent in lensing samples. For example, in the sample compiled here $\sigma
\gsim 1000$ \kms \, are only about twice as numerous as $\sigma \gsim 1500$ \kms
\, systems, suggesting the operation of a mechanism that hinders the
observability of lensed features in low-$\sigma$ clusters. Our discussion above
hints at one possibility: out of all low-mass clusters only those with very
massive central galaxies may display easily identifiable lensed features. In
these systems, tangential arcs are long and thin and appear at large distances
from the center, thereby increasing their visibility to surveys looking
preferentially for images with large length-to-width ratio. Without massive
central galaxies the lensing power of low-mass clusters would be substantially
reduced; arcs would occur closer to the center (where they are hard to
distinguish from a central galaxy), would be significantly magnified in both
radial and tangential directions (see Figure 5 and the discussion by Williams \&
Lewis 1998), and may thus readily have escaped detection as ``giant arcs''.

\section{Summary}

We compare the lensing properties of galaxy clusters with the predictions of
cluster models based on the NFW density profile. We find that clusters are in
general more powerful lenses than NFW halos of similar velocity dispersion. The
magnitude of the discrepancy is small for the most massive clusters in our
sample, $\sigma\sim 1500-2000$ \kms, but increases towards lower cluster
velocity dispersions. NFW lenses also yield large radial magnifications at the
tangential arc location, at odds with observations which indicate that
tangential arc widths are of order of (or perhaps thinner than) the typical
angular size of possible galaxy sources. We use a simple analysis to show that
the data are best reproduced by mass models where the inner slope of the
projected cluster density profile steepens significantly with decreasing
$\sigma$. Agreement with the data requires the effective core lensing potential
to be steeper than isothermal in $\sigma\sim 1000$ \kms clusters, but shallower
than isothermal in the most massive clusters in the sample ($\sigma
\sim 1500$-$2000$ \kms).

We interpret the disagreement between NFW models and lensing observations as
signaling the contribution to the cluster lensing potential of significant
amounts of substructure and of massive central galaxies. Provided that central
galaxy mass correlates only weakly with $\sigma$, its contribution to lensing is
more important in less massive clusters, reproducing the observed trends. We use
N-body simulations to calibrate the effects of substructure, and estimate that
central galaxies as massive as $M_g\approx 3 \times 10^{12} h^{-1} M_{\odot}$
are needed to reconcile NFW halo models with observations. This is much larger
than estimates of the stellar mass of the galaxy; agreement between lensing data
and NFW halo models requires that the central galaxy be surrounded by a dark
matter halo which, within the arc radius, contains almost ten times as much mass
as associated with stars. Lower galaxy masses may be acceptable if, as suggested
by recent N-body experiments, the NFW model systematically underestimates the
central concentration of dark matter halos (Moore et al 1998).

Qualitatively, this conclusion applies to all dark halo models where the inner
slope within the Einstein radius is approximately independent of mass, although
the quantitative estimates presented above are strictly valid only for NFW
models and for the $\Lambda$CDM model we explore. Quantitative estimates,
however, are quite insensitive to the values of the cosmological parameters,
varying only by $10$-$20\%$ when $\Omega_0$ is allowed to vary between $0.2$ and
$1$ (in open and flat geometries).

A crucial ingredient of this interpretation is that the less massive the cluster
the more conspicuous the lensing role played by the central galaxy. The role of
a central galaxy in modifying the cluster's inner profile can in principle be
tested through observations of radial/tangential arc systems. Our modeling
predicts that the redshifts of the radial and tangential arcs must be similar in
MS0440, MS2137, and AC114, but that the radial arc source is far behind the
tangential arc source in A370. Radial arc redshifts are therefore sensitive
tests of our model predictions. These observations are within reach of the
$8$-$10$m class telescopes coming into operation, so we should be able to assess
the validity of the modeling we propose here very soon indeed.

\acknowledgements
This work has been supported in part by the National Science and Engineering
Research Council of Canada. JFN acknowledges useful discussions with Mike Hudson,
Greg Fahlman, and Ian Smail.

\vfill\eject

\begin{table*}
\begin{center}
{Table 1:  Tangential Arcs in Clusters}
\vskip0.1in
\begin{tabular}{lrrrrrrcc}
\tableline
\tableline
Cluster & 
$z_l$ &
$\sigma_v$ &
Arc & $z_s$ &
$\theta$ &
$\delta\theta_t$ &
Telescope &
References \\
		 & 
		 &
[km/s]		 &
		 &
		 &
[asec]		 &
[asec]		 &
		 &
		\\
\tableline
A370        & 0.373 & 1367 & A0      & 0.724 & 10   &    0.54 & WFC1     & 1, 2, 3, 4, 5    \hfil\cr
A963        & 0.206 & 1100 &         & 0.711 & 18.5 & $<$0.50 & UH 2.2m  & 1, 6             \hfil\cr
A1689       & 0.181 & 1989 &         &   -   & 46.5 &      -  &  -       & 1                \hfil\cr
A2124       & 0.066 &  878 &         & 0.573 & 27   &    0.30 & KeckII   & 7                \hfil\cr
A2163       & 0.203 & 1680 &         & 0.728 & 15.5 &      -  &  -       & 1                \hfil\cr
A2218       & 0.171 & 1405 & $\#$359 & 0.702 & 21.2 &    0.47 & WFPC2    & 1, 5, 8, 9       \hfil\cr
A2280       & 0.326 &  948 &         &   -   & 13.9 & $<$0.45 & UH 2.2m  & 1, 10            \hfil\cr
A2390       & 0.228 & 1093 & H1      & 0.913 & 38   &    0.65 & CFHT     & 1, 11, 12, 13    \hfil\cr
A2744       & 0.308 & 1950 &         &   -   & 21.4 &      -  &  -       & 1                \hfil\cr
S295        & 0.299 &  907 &         & 0.93  & 25.5 & $<$0.57 & ESO 3.6m & 1, 14            \hfil\cr
MS0440+0204 & 0.197 &  872 & 2       & 0.53  & 21.8 &    0.40 & UH, CFHT & 1, 15, 16, 17    \hfil\cr
            &       &      & 3       &  ''   & 22.8 & $<$0.25 &  ''      & ''               \hfil\cr
MS0451-0305 & 0.539 & 1371 & A1      &   -   & 32   &    0.45 & UH, CFHT & 1, 16            \hfil\cr
            &       &      & A2      &   -   & 22   &    0.45 &  ''      & ''               \hfil\cr
MS1006+1201 & 0.221 &  906 & A2      &   -   & 27   &    0.60 & UH, CFHT & 1, 16, 18        \hfil\cr
            &       &      & A3      &   -   & 18   &    0.40 &  ''      & ''               \hfil\cr
MS1008-1224 & 0.306 & 1054 & 1       &   -   & 45   &    0.40 & UH, CFHT & 1, 16, 18        \hfil\cr
            &       &      & 2       &   -   & 53   &    0.30 &  ''      & ''               \hfil\cr
MS1358+6245 & 0.329 &  987 &         & 4.92  & 23   & $<$0.40 & UH, CFHT & 1, 16, 19        \hfil\cr
MS1455+2232 & 0.257 & 1133 & 1       &   -   & 21   &    0.65 & UH, CFHT & 1, 16, 18        \hfil\cr
AC114       & 0.310 & 1649 & A0      & 0.639 & 62.4 &    0.78 & WFC1     & 1, 5, 20         \hfil\cr
            &       &      & S1/S2   &   -   & 13.5 &    0.63 & ''       & 1, 5, 21         \hfil\cr
Cl0016+1609 & 0.545 & 1234 & 1       &   -   & 25   & $<$0.10 & WFCP2    & 22               \hfil\cr
Cl0024+1654 & 0.391 & 1339 & C       & 1.7   & 34.6 &    0.78 & WFC1     & 1, 5, 23, 24     \hfil\cr
Cl0302+1658 & 0.423 & 1100 & A1/A1W  & 0.8   & 18.5 &    0.6  & CFHT     & 1, 25            \hfil\cr
            &       &      & A1/A1W  &   -   & 18   & $<$0.25 & UH, CFHT & 1, 16            \hfil\cr
Cl0500-24   & 0.327 & 1152 &         &   -   & 25.9 &      -  & -        & 1, 26            \hfil\cr
MS1621+2640 & 0.427 &  793 & A1      &   -   &  6.9 & $<$0.25 & -        & 1, 16, 27        \hfil\cr
MS2137-2353 & 0.313 &  960 & A0      &   -   & 15.5 &    0.40 & ESO NTT  & 1, 28            \hfil\cr
RX1347-1145 & 0.451 & 1235 & 1       &   -   & 34.2 &    0.38 & STIS     & 1, 29, 30        \hfil\cr
            &       &      & 4       &   -   & 36.3 &    0.25 & ''       & ''               \hfil\cr
\tableline
\end{tabular} 
\end{center} 

$^1$ Wu et al (1998);
$^2$ Soucail et al (1987a);
$^3$ Soucail et al (1987b);
$^4$ Soucail et al (1988);
$^5$ Smail et al (1996);
$^6$ Lavery \& Henry (1988);
$^7$ Blakeslee et al (1999);
$^8$ Le Borgne et al (1992);
$^9$ Pello et al (1992);
$^{10}$ Gioia et al (1995);
$^{11}$ Pello et al (1991);
$^{12}$ Le Borgne et al (1991);
$^{13}$ Mellier (1989);
$^{14}$ Edge et al (1994);
$^{15}$ Gioia et al (1998);
$^{16}$ Luppino et al (1999);
$^{17}$ Luppino et al (1993);
$^{18}$ Le Fevre et al (1994)
$^{19}$ Luppino et al (1991);
$^{20}$ Smail et al (1991);
$^{21}$ Smail et al (1995);
$^{22}$ Lavery (1996);
$^{23}$ Colley et al (1996);
$^{24}$ Koo (1988);
$^{25}$ Mathez et al (1992);
$^{26}$ Giraud (1988);
$^{27}$ Luppino \& Gioia (1992);
$^{28}$ Fort et al (1992);
$^{29}$ Sahu et al (1998);
$^{30}$ Schindler et al (1995)

\end{table*} 

\clearpage

\begin{figure} 
\plotone{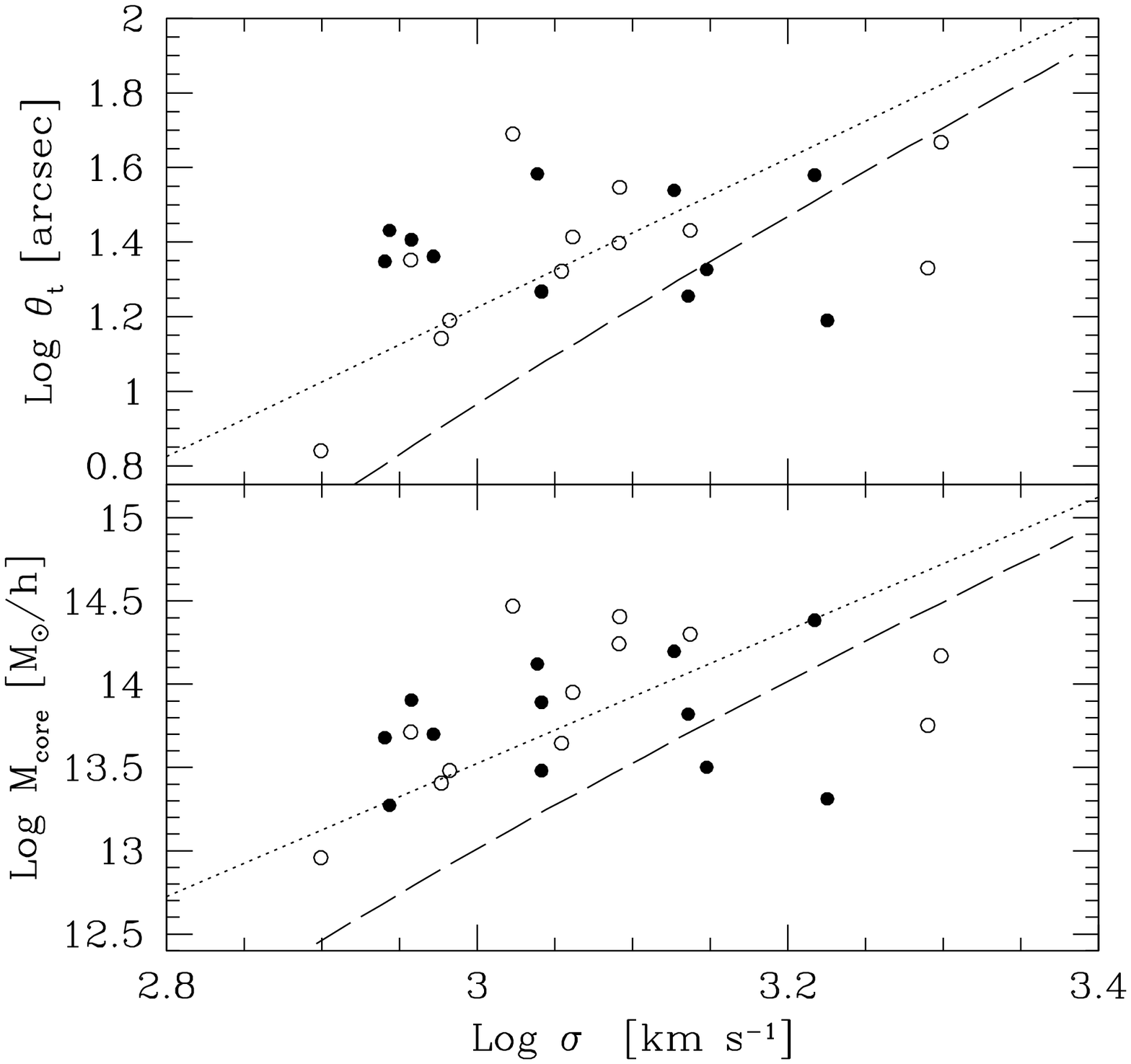}
\figurenum{1} 
\caption{The strong lensing properties of galaxy clusters in our sample as a
function of cluster velocity dispersion, $\sigma$. Arcs without spectroscopic
redshifts are assumed to be at $z_s \sim 1$, and are marked by empty circles.
The dotted and dashed lines are for singular isothermal sphere and NFW cluster
models, respectively, assuming that clusters are located at $z_l=0.3$ and that
the sources are at $z_s=1$.  {\it Top panel:} clustercentric distance of the
tangential arc (the ``Einstein radius'') in arcsec. {\it Bottom panel:} cluster
core mass derived from the top panel assuming circular symmetry and angular
diameter distances appropriate to an Einstein-de Sitter universe.}
\end{figure}   

\begin{figure} 
\plotone{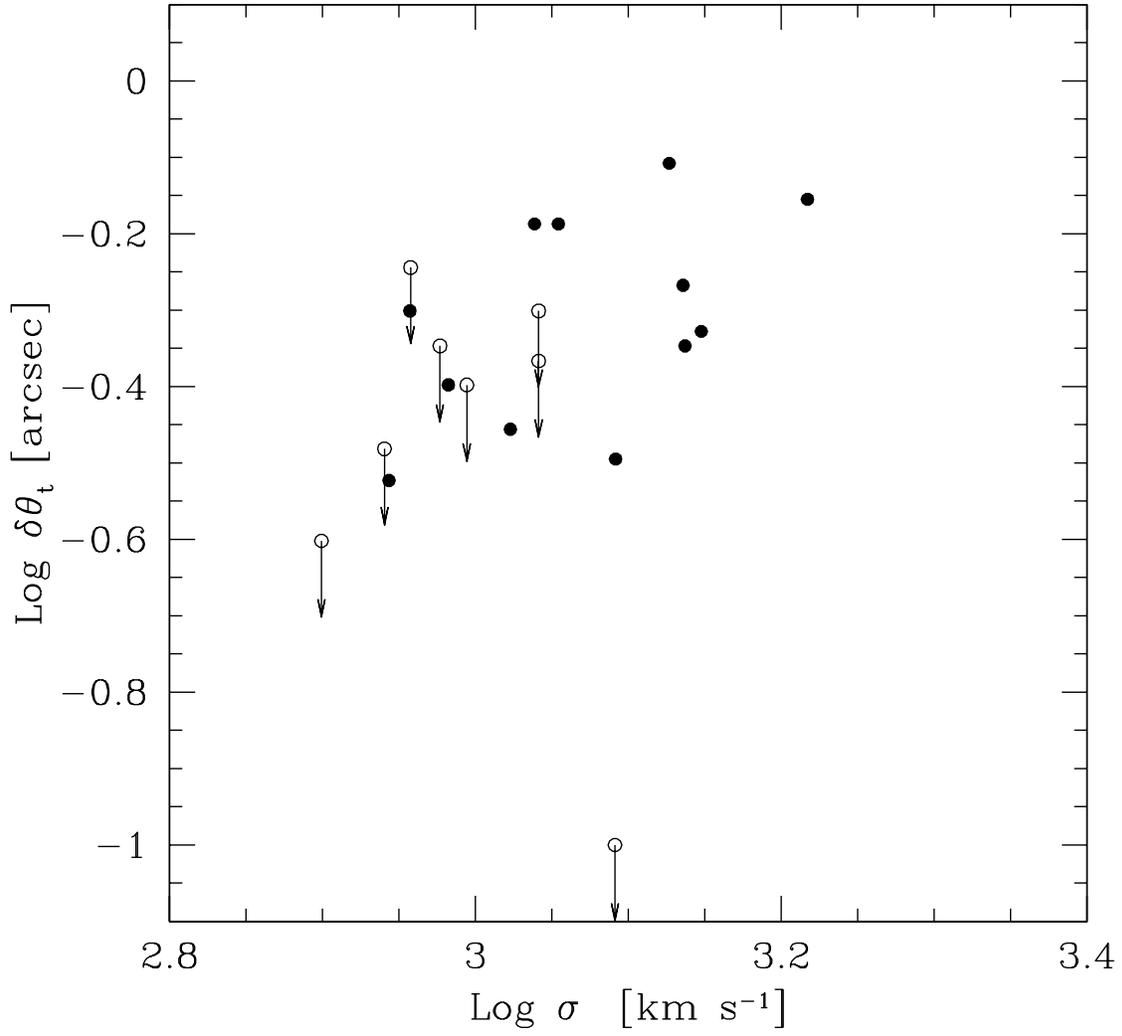}
\figurenum{2} 
\caption{Tangential arc ``half-light'' widths (solid circles) or upper limits based 
on the seeing at the time of observation (open circles).}
\end{figure}   

\begin{figure} 
\plotone{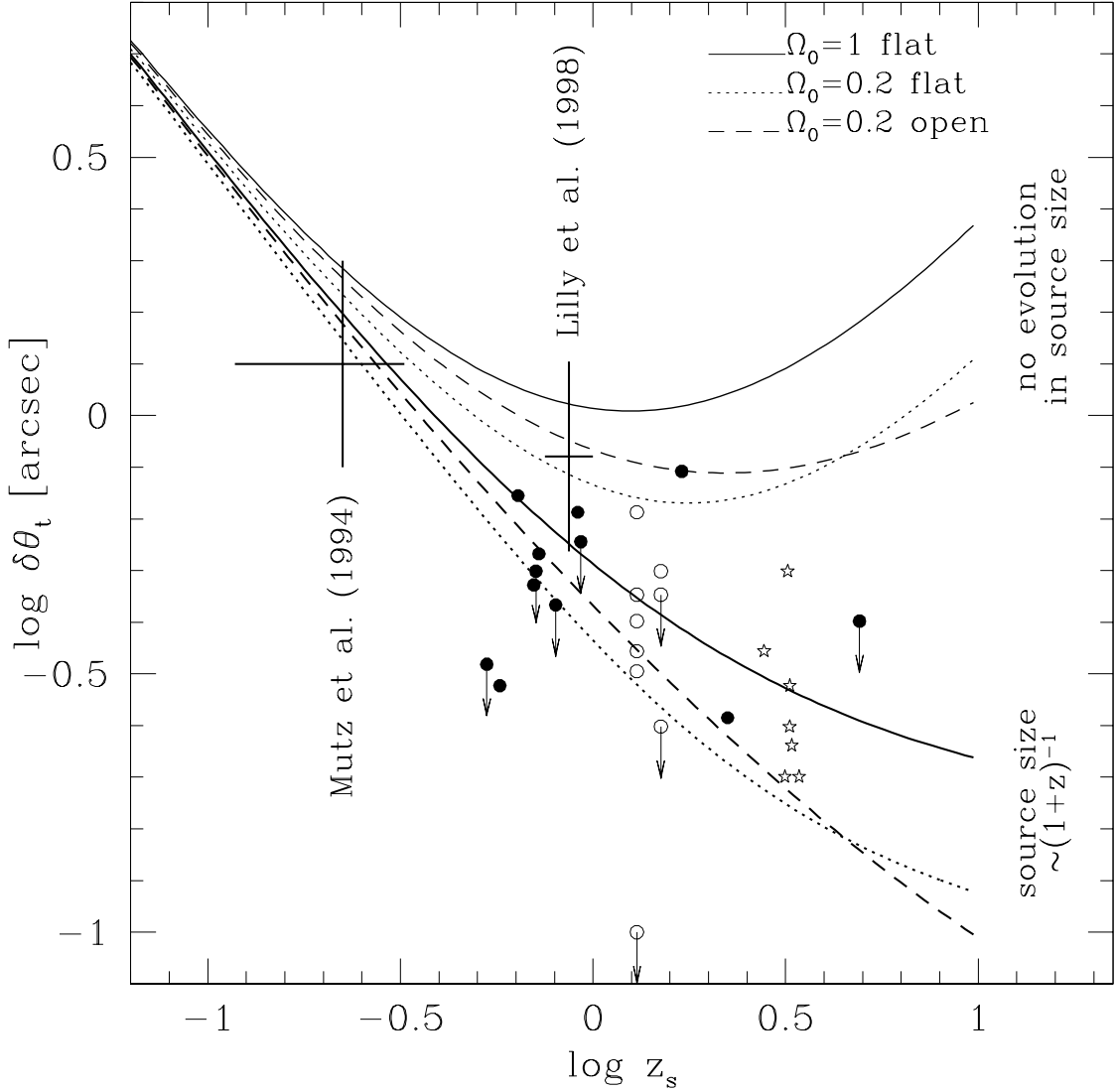}
\figurenum{3} 
\caption{The tangential arc half-widths (filled circles) compared with the intrinsic
angular size of galaxies at different redshifts from the HST MDS survey of Mutz
et al (1994), the CFRS survey of Lilly et al (1998), and the Ly-break galaxies
of Giavalisco et al (1996) (starred symbols). Arcs without redshifts are shown with
open circles. Note that arc widths are comparable (or smaller) than the angular
size of galaxies at comparable redshifts, indicating that the radial
magnification at the tangential critical line is about (or less than) unity.}
\end{figure}   

\begin{figure} 
\plotone{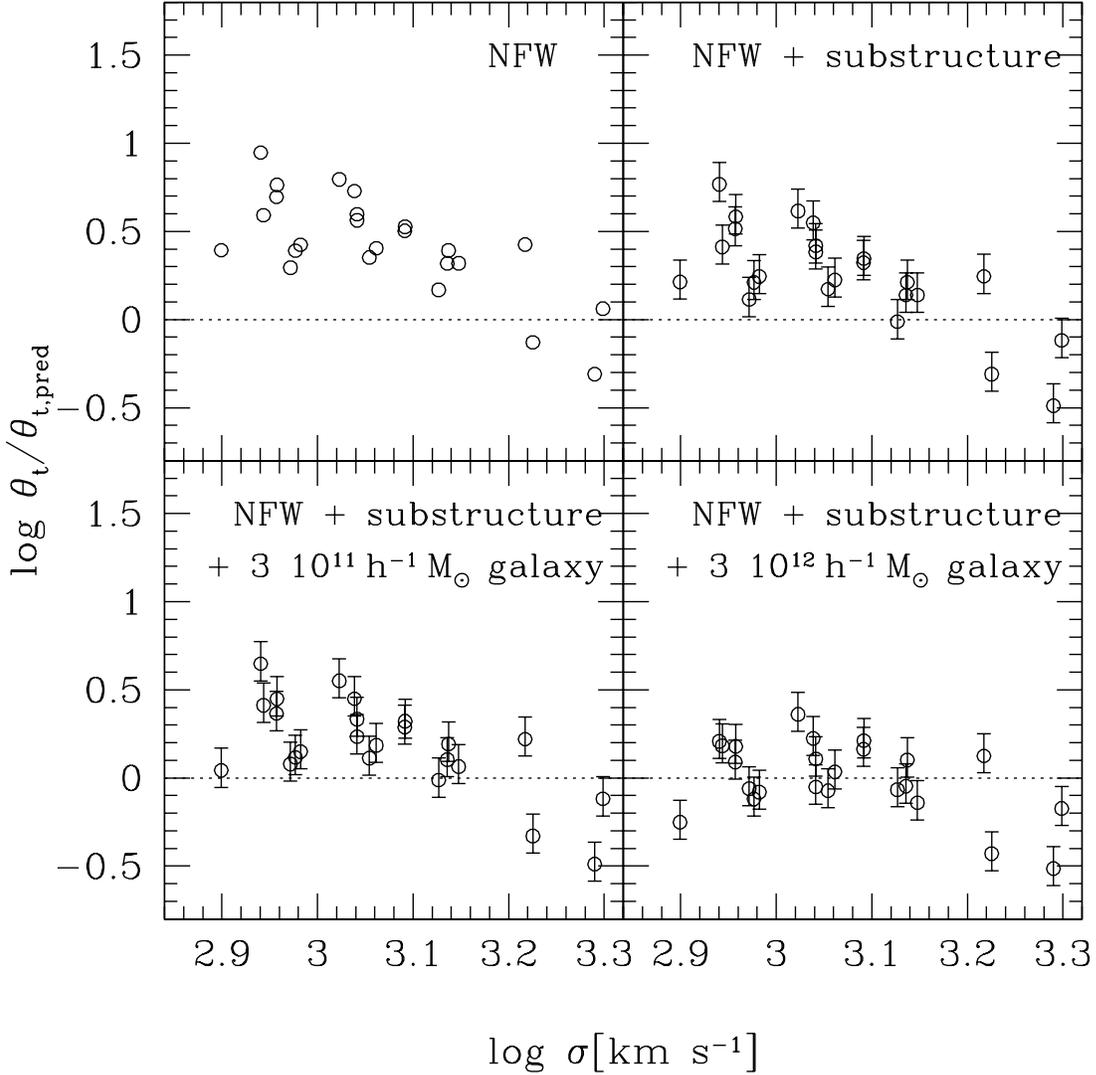}
\figurenum{4} 
\caption{{\it Top left:} Ratio between the observed clustercentric distance of
tangential arcs and the predictions of NFW halo models in an $\Omega_0=0.2$,
$\Lambda=0.8$, $h=0.7$ cluster-normalized cold dark matter dominated universe.
{\it Top right:} as in top left panel, but assuming that the lensing properties
of NFW halos are aided by substructure as described in \S 4.4. The error bars
represent the $1/4$ and $3/4$ quartile range of the substructure $f_t$
parameter, see Figure 6.  {\it Bottom left:} as in top left panel, but assuming
that the lensing properties of NFW halos are aided by the presence of
substructure and a central galaxy with $M_g=3 \times 10^{11}h^{-1}M_\odot$, see
\S 4.5.  {\it Bottom right:} as in bottom left, but with $M_g=3 \times
10^{12}h^{-1}M_\odot$.}
\end{figure}   

\begin{figure} 
\plotone{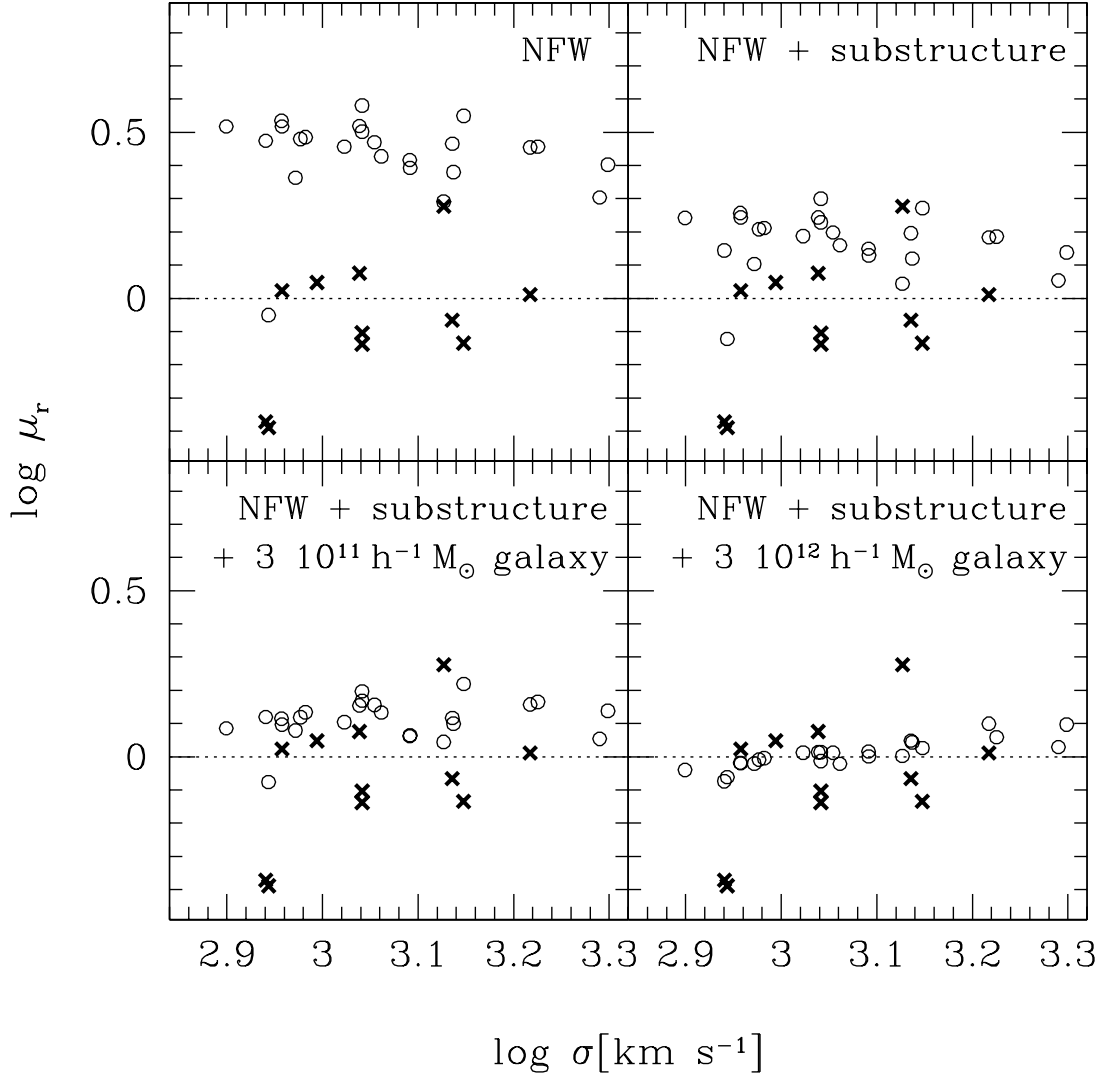}
\figurenum{5} 
\caption{Panels are as in Figure 4, but for the radial magnification at the
tangential critical line, $\mu_r(\theta_t)$ (empty circles). Crosses indicate
magnification estimates based on observed arc widths and redshifts, and the
assumption that the intrinsic angular size of the source is $4.4 \,
(1+z_s)^{-1/2} \, h^{-1}$ kpc.}
\end{figure}   

\begin{figure} 
\plotone{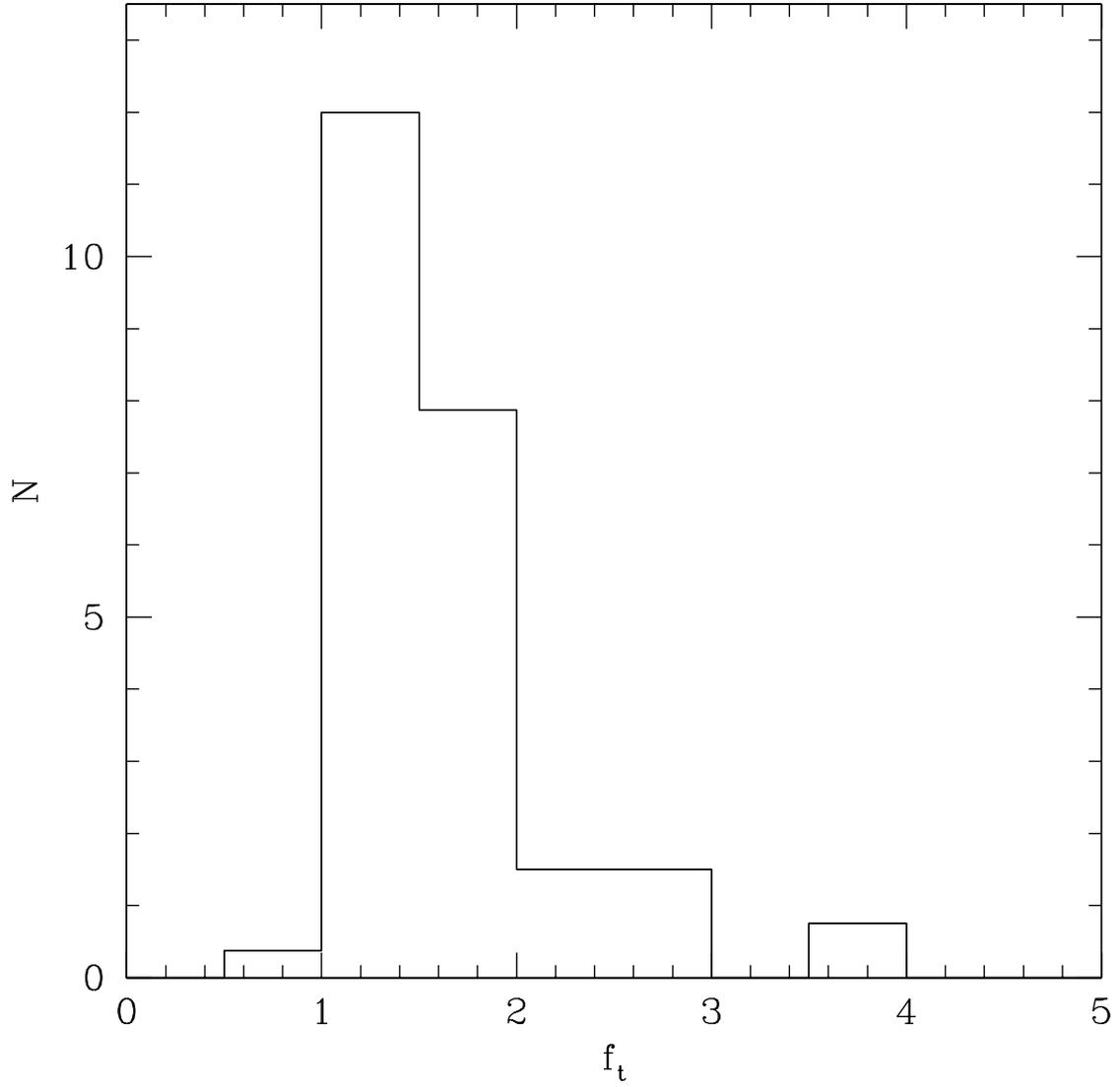}
\figurenum{6} 
\caption{The distribution of substructure parameter, $f_t$, obtained from
N-body simulations and used to estimate ``uncertainties'' in Figures 4 and 5.}
\end{figure}   

\begin{figure} 
\plotone{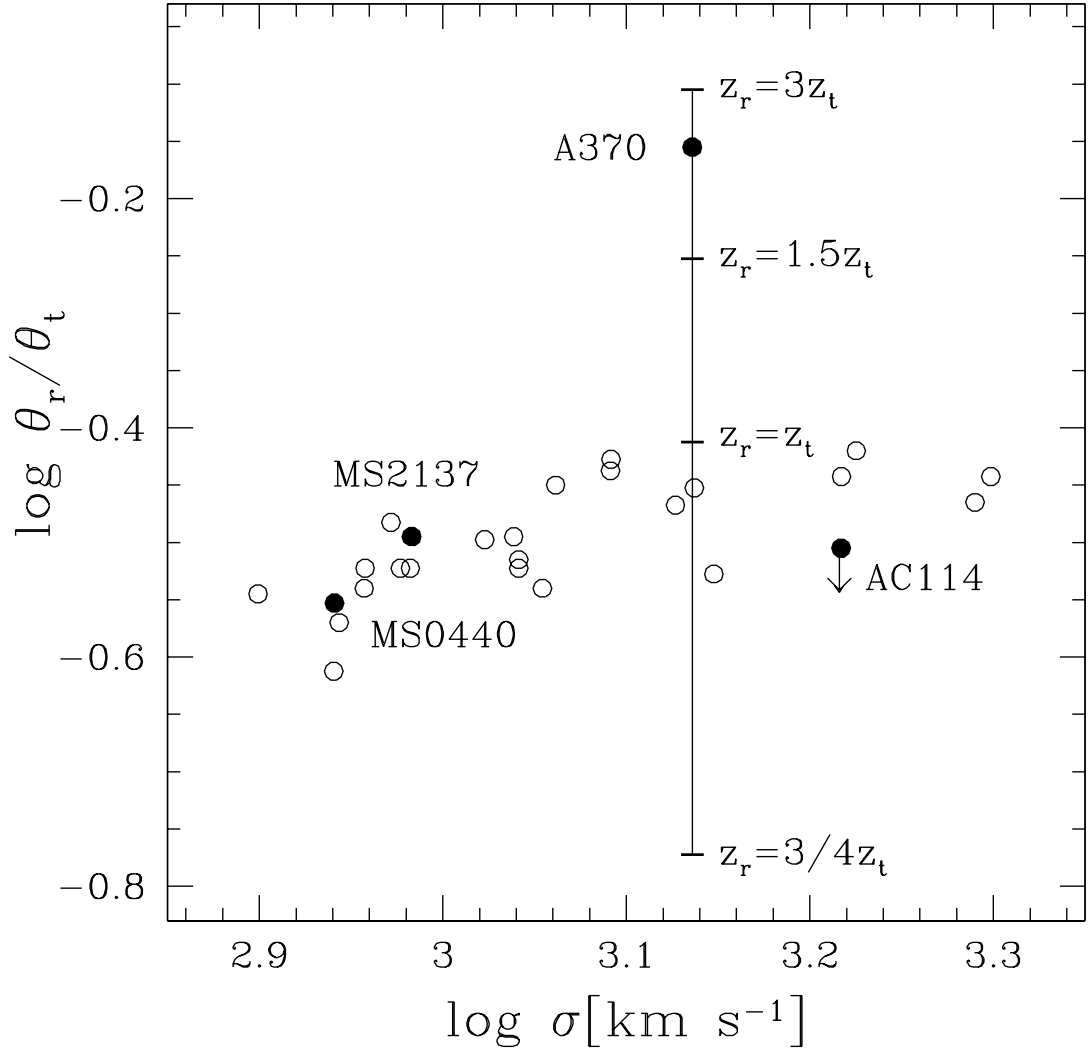}
\figurenum{7} 
\caption{The ratio between clustercentric distance of radial and tangential arcs
for NFW cluster models that include a $M_g=3 \times 10^{11}\, h^{-1} M_\odot$
central galaxy, as described in \S 4.5.  Open circles are the model predictions
as a function of cluster velocity dispersion, assuming that both arcs are at the
same redshift. Tangential arc redshifts are assumed to be $z_s=1$ if
unavailable. Data for MS0440, MS2137 and AC114 are consistent with this
assumption, but A370 deviates strongly from the predicted trend. This may
indicate that the radial arc source is far behind the tangential arc galaxy, at
$z_r\approx 2$-$3 \, z_t$. A spectroscopic redshift determination of this arc
would be useful for assessing the validity of our mass model.}
\end{figure} 

\end{document}